\documentstyle[amsmath,amssymb,graphicx]{article}

\def\be{\begin{eqnarray}}
\def\ee{\end{eqnarray}}
\def\nn{\nonumber}

\def\p{\partial}
\def\tr{{\rm tr}\,}
\def\Tr{{\rm Tr}\,}

\newdimen\linethick  \linethick=0.4pt
\newdimen\hboxitspace    \hboxitspace=5pt
\newdimen\vboxitspace    \vboxitspace=5pt

\def\fr#1{%
\be
\vcenter{
\hrule height\linethick
          \hbox{\vrule width\linethick
                \kern\hboxitspace
                \vbox{\kern\vboxitspace
                      \hbox{$\begin{array}{c}\displaystyle#1
         \end{array}$}%
                      \kern\vboxitspace}%
                \kern\hboxitspace
                \vrule width\linethick}%
          \hrule height\linethick}%
\ee}

\hoffset= - 1.0in         

\title{{\bf Matrix model version of AGT conjecture and generalized Selberg
integrals} \vspace{.2cm}}
\author{{\bf A.Mironov}\footnote{ {\small {\it
Lebedev Physics Institute} and {\it ITEP, Moscow, Russia}};
mironov@itep.ru; mironov@lpi.ru}, {\bf Al.Morozov}\thanks{{\small
{\it ITEP, Moscow, Russia}}; morozov@itep.ru} \ and {\bf
And.Morozov}\thanks{{\small {\it ITEP, Moscow, Russia} and {\it
Physics Department, Moscow State University, Moscow, Russia}};
andrey.morozov@itep.ru}\date{ }}

\begin{document}

\maketitle

\vspace{-6.0cm}

\begin{center}
\hfill FIAN/TD-04/10\\
\hfill ITEP/TH-14/10\\
\end{center}

\vspace{4cm}

\begin{abstract}
Operator product expansion (OPE)
of two operators in two-dimensional conformal field
theory includes a sum over Virasoro descendants of other
operator with universal coefficients, dictated exclusively by
properties of the Virasoro algebra and independent of choice of
the particular conformal model. In the free field model, these coefficients
arise only with a special "conservation" relation imposed on the
three dimensions of the operators involved in OPE.
We demonstrate that the coefficients for the three
unconstrained dimensions arise in the free field formalism when
additional Dotsenko-Fateev integrals are inserted between the
positions of the two original operators in the product. If such
coefficients are combined to form an $n$-point conformal block on
Riemann sphere, one reproduces the earlier conjectured
$\beta$-ensemble representation of conformal blocks, thus proving
this (matrix model) version of the celebrated AGT relation. The
statement can also be regarded as a relation between the $3j$-symbols of
the Virasoro algebra and the slightly generalized Selberg integrals
$I_Y$, associated with arbitrary Young diagrams. The conformal blocks
are multilinear combinations of such integrals and the remaining
part of the original AGT conjecture relates them to the Nekrasov
functions which have exactly the same structure.
\end{abstract}

\bigskip

\bigskip

\section{Introduction}

The AGT conjecture \cite{AGT} unifies and identifies
a number of different domains in modern theory,
what makes it a very interesting and promising subject,
attracting a lot of attention \cite{AGTfirst}-\cite{IO}.
In its original form the AGT conjecture relates the conformal
blocks in $2d$ conformal field theory (CFT) \cite{CFT1,CFT2} and the
Nekrasov functions \cite{NF},
obtained by expansion of the LNS multiple contour integrals \cite{LNS}.
In this form it is now proved only in three cases:
in the limit of large central charge $c$  \cite{MMc}, when the conformal
blocks and the Nekrasov functions reduce to (generic) hypergeometric
series; in the case of special value of one of the external dimensions
\cite{MMnf,MMu3}, when they are also hypergeometric series (however, different from
the first case); and in the case of a 1-point toric function \cite{FLit2},
when one can use the powerful Zamolodchikov recurrent relation
\cite{Zarec,Pog}.
Following earlier considerations in \cite{MMnf,DVagt,Ito,Egu,Wilma,MMSh1}
in \cite{MMSh2} a simpler version of the AGT relation was suggested,
identifying conformal blocks with the Dotsenko-Fateev $\beta$-ensemble
integral \cite{DF,MMMconf,UFN3,DVagt},
which can be considered as a new avatar of the old proposal in
the free field approach to CFT \cite{DF,GMMOS}
and, at the same time, as a concrete application of the more recent theory
of Dijkgraaf-Vafa (DV) phases of matrix models \cite{DVpha}.
Ref.\cite{MMSh2} contains absolutely explicit formulas
for generic conformal blocks, made from the free field correlators
with screening integral insertions, analytically continued in the
number of screenings.
The only problem is that these formulas are very tedious to derive
and their meaning from the point of view of representation theory
of the Virasoro algebra, an underlying algebra for
the standard construction of the
conformal blocks in \cite{CFT1,CFT2}, remains no less obscure than in the
original AGT relation of \cite{AGT}.

In this paper we provide a similar, but conceptually different
derivation of the same formulas of \cite{MMSh1,MMSh2},
which involves nothing but the Virasoro representation theory and
by now the elementary Selberg integrals \cite{Selint}.
Calculations remain tedious but now they are conceptually clear
and straightforward.
We give only basic examples, but a full {\it constructive} proof
can definitely be worked out in this way, and in this sense one may say
that the simplified version \cite{MMSh2}
of the AGT conjecture \cite{AGT} is now practically established.
In fact, the {\it conceptual} proof is readily available and is given
in s.11.1 in the conclusion.
It can be further promoted to a straightforward proof of the original
AGT conjecture: after the recent progress in \cite{IO}
there remain just a few combinatorial details to fix.
In this sense {\bf the program to prove the AGT conjecture
through the technique
of the Dotsenko-Fateev (matrix model like) integrals, which was formulated in
\cite{MMnf} and \cite{DVagt}, is nearly completed}.
Still it would be interesting to work out some other proof,
establishing an explicit relation between the Dotsenko-Fateev
and LNS integrals, perhaps, making use of the duality between Gaussian
and Kontsevich models:
this is, however, only mentioned in s.11.2 in the Conclusion
and remains beyond the scope of the present paper.

\bigskip

In this paper we consider the triple functions, the coefficients in
the operator product expansion involving a sum over Virasoro
descendants, in the free field formalism.
We begin with
the case of the single free field, i.e. with the pure Virasoro
chiral algebra
(in the AGT terminology this corresponds to the $U(2)^{\otimes (n-3)}$ case),
the extension to $k$ fields, the $W_{k+1}$ chiral algebra
and the $U(k+1)$ quiver is straightforward.
In terms of the Virasoro primary, elements of the Verma modules
look like $:L_{-Y}V_\Delta:$ and the operator product expansion in the
corresponding {\it chiral algebra} is
\be
:L_{-Y_1}V_{\Delta_1}(0):\
:L_{-Y_2}V_{\Delta_2}(q):\ =
\sum_\Delta
q^{\Delta - \Delta_1 - \Delta_2 - |Y_1|-|Y_2|} S_{\Delta_1\Delta_2}^{\Delta}
\sum_{Y} q^{|Y|}
C_{\Delta_1,Y_1;\ \Delta_2,Y_2}^{\Delta,Y} :L_{-Y}V_\Delta(0):
\label{OPE}
\ee
Here $V_\Delta$ denotes the Virasoro primary with dimension $\Delta$:
$L_n V_\Delta = 0$ for $n>0$,
$L_0 V_\Delta = \Delta V_\Delta$, and $L_{-Y} = \ldots L_{-n_2}L_{-n_1}$,
denotes the "negative" (raising) Virasoro operator, labeled by the Young diagram
$Y=\{n_1\geq n_2\geq \ldots\}$.
The coefficients $S_{\Delta_1\Delta_2}^\Delta$
depend on choice of the conformal model. In particular,
their values, directly provided by the free field formalism below,
are usually referred to as the Liouville model structure constants
\cite{DOZZ},
we do not consider other choices in this paper.
In contrast to $S_{\Delta_1\Delta_2}^\Delta$, the coefficients
$C_{\Delta_1,Y_1;\ \Delta_2,Y_2}^{\Delta,Y}$
are universal, depend only on the properties of the Virasoro algebra,
and these are the quantities we are going to investigate.
Moreover, we further restrict our consideration to the case of $Y_2=0$,
this is enough to reproduce the spherical 4-point conformal blocks, studied
in \cite{AGT,AlMo2,MMSh2}.

The coefficients $C$ can be straightforwardly found by standard
CFT methods, see \cite{MMMM} for a detailed review. Coming back to
free fields, they provide an alternative derivation, somewhat
simpler and more transparent, see \cite{MMMM} and s.\ref{fref}.
The only problem is that in the free field model there is a
"conservation law": the primaries are represented as $V_\Delta =
e^{\alpha\phi}$ with\footnote{ In terms of parametrization from
the Appendix of \cite{MMu3}, normalization conventions for free fields
in the present paper are as follows: $k_1=1/2$, $k_2= 2$, $p = 1/\lambda
= 1$, where \be
\phi(z)\phi(0) = k_2\log z, \nn \\
T = \frac{1}{2k_2}(\p\phi)^2 + k_1Q\p^2\phi, \nn \\
c = 1-12k_1^2k_2Q^2, \nn \\
V_\alpha = e^{p\alpha\phi}, \ \ \ \
V_{\alpha_1}(z) V_{\alpha_2}(0) = z^{p^2k_2\alpha_1\alpha_2}, \nn \\
\Delta(V_\alpha) = \frac{p^2k_2}{2}\alpha(\alpha-Q), \ \ \ \ \sum_i
\alpha_i = 2k_1Q/p \nn \ee Then \be
:L_{-1}^2V_\alpha:\ = \p^2V_\alpha =
:\left(p^2\alpha^2(\p\phi)^2 + p\alpha\p^2\phi\right)V_\alpha :\ ,\nn\\
:L_{-2}V_\alpha:\ = :\left(\frac{1}{2k_2}(\p\phi)^2 +
(k_1Q+p\alpha)\p^2\phi\right)V_\alpha: \nn \ee }
\be \Delta = \alpha(Q-\alpha),\ \ \ \ c =
1-6Q^2, \label{Delal}
\ee
and in the sum at the r.h.s. of
(\ref{OPE}) \be \alpha = \alpha_1+\alpha_2 \ee Thus, only a restricted
set of the triple functions $C$, namely, $C_{\alpha_1,Y_1;\
\alpha_2,Y_2}^{\alpha_1+\alpha_2,Y}$ can be defined in this model
(from now on we label these functions with $\alpha$- rather than
$\Delta$-parameters). It is a long-standing problem in CFT, how the
free field formalism can be used to obtain arbitrary
$C_{\alpha_1,Y_1;\ \alpha_2,Y_2}^{\alpha,Y}$ with $\alpha\neq
\alpha_1+\alpha_2$. The results of \cite{MMSh1,MMSh2} imply that the
operator product \be \boxed{ :L_{-Y_1}e^{\alpha_1\phi(0)}:\
:L_{-Y_2}e^{\alpha_2\phi(q)}:\ \left(\int_0^q
:e^{b\phi(z)}:\,dz\right)^N = \tilde
S_{\alpha_1\alpha_2}^{\alpha_1+\alpha_2+bN} \sum_Y q^{|Y|}\tilde
C_{\alpha_1,Y_1;\ \alpha_2,Y_2}^{\alpha_1+\alpha_2+bN,Y}
:L_{-Y}e^{(\alpha_1+\alpha_2+bN)\phi(0)}: } \label{OPEfs} \ee where
$b$ is the Dotsenko-Fateev screening charge, i.e. $Q=b-1/b$, has
exactly the same expansion coefficients as OPE, \be \boxed{
\tilde C_{\alpha_1\alpha_2}^{\alpha,Y} = C_{\alpha_1\alpha_2}^{\alpha,Y},
}
\label{SSCC}
\ee
and we demonstrate
below that this is indeed the case.
Eq.(\ref{SSCC}) is the main claim of the present paper, supported by
a number of examples.
In other words,
the r.h.s. of eq.(\ref{OPEfs}) is identically the same as (\ref{OPE})
provided the structure constants $C$ and $\tilde C$
are related by a change of variables (\ref{Delal}) and additionally
\be
\alpha = \alpha_1+\alpha_2+bN
\ee
Thus, (\ref{OPEfs}) resolves the above mentioned problem in the sense of
analytical continuation: the coefficients
$\tilde C_{\alpha_1,Y_1;\ \alpha_2,Y_2}^{\alpha,Y}$
are rational functions of $\alpha$ and they are fully defined by
their values at discrete points $\alpha = \alpha_1+\alpha_2+bN$.

If the original two fields are primaries, $Y_1=Y_2=\emptyset$,
then eq.(\ref{OPEfs}) is derived in three steps.

A) First, one uses the basic free field relation,
\be
:e^{\alpha_1\phi(0)}:\ :e^{\alpha_2\phi(q)}:\ \prod_{i=1}^N :e^{b\phi(z_i)}:\ =
\left\{q^{2\alpha_1\alpha_2} \prod_{i<j}^N (z_i-z_j)^{2b^2}
\prod_{i=1}^N z_i^{2b\alpha_1}(q-z_i)^{2b\alpha_2}\right\}
\ : e^{\alpha_1\phi(0) + \alpha_2\phi(q) + b\sum_i\phi(z_i)}:\
\label{frefope}
\ee
and then expands the exponential in powers of $q$ and $z_i$:
\be
:e^{\alpha_1\phi(0) + \alpha_2\phi(q) + b\sum_i\phi(z_i)}:\ =
\sum_{Y,Y'}  q^{|Y|-|Y'|}
H_{Y'Y} z^{Y'} :L_{-Y}e^{(\alpha_1+\alpha_2+Nb)\phi(0)}:
\ee
Here $z^{Y'} = \prod_i z_i^{n_i}$
for a Young diagram $Y' = \{n_1\geq n_2\geq\ldots\}$ and the sum
goes over all pairs of Young diagrams with $|Y|\geq |Y'|$.
At this step, one evaluates the $z$-independent coefficients $H_{YY'}$,
as functions of $\alpha_1,\alpha_2$ and $N$.

B) Next, one takes the integrals over $z_i$,
\be
I_{Y'} =
\prod_{i=1}^N\int_0^q dz_i \left\{z^{Y'} \prod_{i<j}^N (z_i-z_j)^{2b^2}
\prod_{i=1}^N z_i^{2b\alpha_1}(q-z_i)^{2b\alpha_2}\right\}
\label{Selin1}
\ee
For the single-line Young diagrams $Y' = [1^n]$ these are
the well known Selberg integrals, which generalize the Euler
$B$-function and are equal
(after the standard analytical continuation from integer powers
in the integrand) to the ratio of $Gamma$-factors.
For generic $Y'$, the integrals generalize the Selberg integrals
producing extra non-factorizable polynomial
factors, which can be explicitly evaluated.
Being polynomial, they do not complicate the analytical
continuation.

C) Combining the results of steps A and B, one gets the structure
\be
\tilde C_{\alpha_1;\alpha_2}^{a,Y} = \left.\sum_{|Y'|\leq |Y|}
H_{YY'}I_{Y'}\right|_{q=1}
\ee
in the form of a finite sums over Young diagrams.

D) The last step is to compare the $\tilde C$ with the known
expressions for the conformal theory structure constants $C$
(the $3j$-symbols of the Virasoro algebra), transformed with the help
of (\ref{Delal}).

For non-trivial $Y_1$ and $Y_2$ the calculation goes the same way,
with additional powers of $z_i^{-1}$ and $(z_i-q)^{-1}$ emerging
in the integrand.

\bigskip

In this letter we provide in full detail a sample calculation
of this kind for the two simplest cases of
$\{Y_1,Y_2,Y\} = \{[0],[0],[1]\}$, $\{[0],[0],[2]\}$ and
$\{[0],[0],[11]\}$.
It is enough to demonstrate the principle and can be straightforwardly
computerized to provide more examples.
There is small doubt that all such examples would confirm the
relation, which at the moment looks like a non-trivial
statement, identifying the $3j$-symbols of the Virasoro algebra
with linear combinations of the generalized Selberg integrals
$I_{Y}$.

\section{The free field formulas for $C_{\alpha_1\alpha_2}^{\alpha_1+\alpha_2}$
\label{fref}}

Evaluation of the operator product coefficients in the free field model
is considered in detail in \cite{MMMM}.
The simplest example is:
\be
:e^{\alpha_1\phi(0)}:\ :e^{\alpha_2\phi(q)}:\ =\
q^{2\alpha_1\alpha_2}\ :e^{\alpha_1\phi(0)+\alpha_2\phi(q)}:\ =\nn \\ =
q^{2\alpha_1\alpha_2}\ :\left(1 + q\alpha_2\p\phi(0)
+ \frac{q^2}{2}\Big(\alpha_2\p^2\phi(0) + \alpha_2^2(\p\phi(0))^2\Big)
+ \ldots\right)\,e^{(\alpha_1+\alpha_2)\phi(0)}:
\label{frefOPE}
\ee
Now, using
\be
:L_{-1}e^{\alpha\phi}:\ =\ : \alpha\p\phi e^{\alpha\phi}: \nn \\
:L_{-1}^2e^{\alpha\phi}:\ =\ : \Big(\alpha\p^2\phi + \alpha^2(\p\phi)^2\Big)
e^{\alpha\phi}: \nn \\
:L_{-2}e^{\alpha\phi}:\ =\ : \left(\frac{1}{4}(\p\phi)^2 +
\Big({1\over 2}Q+\alpha\Big)\p^2\phi\right)
e^{\alpha\phi}:\nn
\ee
from
\be
:e^{\alpha_1\phi(0)}:\ :e^{\alpha_2\phi(q)}:\
= \sum_Y q^{|Y|} C_{\alpha_1\alpha_2}^{\alpha_1+\alpha_2,L_{-Y}}
\ :L_{-Y}e^{(\alpha_1+\alpha_2)\phi(0)}:
\ee
one obtains
\fr{
C_{\alpha_1\alpha_2}^{\alpha_1+\alpha_2,\,L_{-1}} = \frac{\alpha_2}{\alpha_1+\alpha_2}
\\
\\
C_{\alpha_1\alpha_2}^{\alpha_1+\alpha_2,\,L_{-1}^2} =\displaystyle{
\frac{4\alpha_2^2+\alpha_2\left(2\alpha_2Q+4\alpha_1\alpha_2-1\right)}
{2\left(4\left(\alpha_1+\alpha_2\right)^2+2Q\left(\alpha_1+\alpha_2\right)-1\right)
\left(\alpha_1+\alpha_2\right)}}
\\
\\
C_{\alpha_1\alpha_2}^{\alpha_1+\alpha_2,\,L_{-2}} =
\displaystyle{\frac{2\alpha_1\alpha_2}
{4\left(\alpha_1+\alpha_2\right)^2+2Q\left(\alpha_1+\alpha_2\right)-1}}
\label{Cfref001}}

\section{The CFT formulas for $C_{\alpha_1\alpha_2}^{\,\alpha}$}

According to  \cite[eq.(5.16)]{CFT2}, for three generic dimensions one has,
instead of (\ref{Cfref001}),
\be
C_{\alpha_1\alpha_2}^{\alpha,L_{-1}} = \frac{\Delta_2+\Delta-\Delta_1}{2\Delta}
\label{Cfrefsc001}
\ee
Similarly,
\be
C_{\alpha_1\alpha_2}^{\alpha,L_{-1}^2} =
\frac{8\Delta^3+c\Delta^2+16\Delta^2\Delta_2+2\Delta\Delta_2c-16\Delta^2\Delta_1-2\Delta\Delta_1c-4\Delta^2+c\Delta+
8\Delta_2^2\Delta}{4\Delta(16\Delta^2-10\Delta+c+2c\Delta)}
+\nn\\
+\frac{\Delta_2^2c-16\Delta\Delta_1\Delta_2-2\Delta_1\Delta_2c-16\Delta\Delta_2+\Delta_2c+8\Delta_1^2\Delta+\Delta_1^2c+
4\Delta\Delta_1-\Delta_1c}{4\Delta(16\Delta^2-10\Delta+c+2c\Delta)}
\ee
\be
C_{\alpha_1\alpha_2}^{\alpha,L_{-2}} =
\frac{\Delta^2+2\Delta\Delta_2+2\Delta\Delta_1-\Delta-3\Delta_2^2+6\Delta_1\Delta_2+\Delta_2-3\Delta_1^2+\Delta_1}
{16\Delta^2-10\Delta+c+2c\Delta}
\ee
and so on.

Only in the case of the conservation law condition,
\be
\alpha=\alpha_1+\alpha_2
\label{conserv}
\ee
these expressions are reproduced by the free-field formula (\ref{Cfref001}).

\section{Operator product with screening insertions}

In order to relax the $U(1)$ conservation law (\ref{conserv})
we insert the Dotsenko-Fateev screening charges into the l.h.s. of (\ref{frefOPE}).
Then, in the integrand one has
$$
:e^{\alpha_1\phi(0)}:\ :e^{\alpha_2\phi(q)}:\ \prod_{i=1}^N :e^{b\phi(z_i)}:\ =
q^{2\alpha_1\alpha_2} \prod_{i<j}^N (z_i-z_j)^{2b^2}
\prod_{i=1}^N z_i^{2b\alpha_1}(q-z_i)^{2b\alpha_2}
\ : e^{\alpha_1\phi(0) + \alpha_2\phi(q) + b\sum_i\phi(z_i)}:\ =
$$ \vspace{-0.5cm}
\be\nn
= q^{2\alpha_1\alpha_2} \prod\limits_{i<j}^N (z_i-z_j)^{2b^2}
\prod\limits_{i=1}^N z_i^{2b\alpha_1}(q-z_i)^{2b\alpha_2}\:\Big(1 +
\underline{\Big(\alpha_2 q + b\sum_{i=1}^N z_i\Big) \p\phi(0)} +
\\ \vspace{-0.5cm}
+ \underline{\underline{
\left(\alpha_2 q + b\sum\limits_{i=1}^N z_i\right)^2 \frac{\left(\partial\phi(0)\right)^2}{2!} +
\left(\alpha_2 q^2 + b \sum\limits_{i=1}^N z_i^2\right)\frac{\partial^2\phi(0)}{2!}
}}+
\nn \\ \nn
+ \underline{\underline{\underline{
\left(\alpha_2 q + b\sum\limits_{i=1}^N z_i\right)^3
\frac{\left(\partial\phi(0)\right)^3}{3!}+
3\cdot\left(\alpha_2 q + b\sum\limits_{i=1}^N z_i\right)
\left(\alpha_2 q^2 + b\sum\limits_{i=1}^N z^2_i\right)\frac{\partial\phi(0)\partial^2\phi(0)}{3!}+
}}}\\ \nn \left. \underline{\underline{\underline{
+\left(\alpha_2 q^3 + b\sum\limits_{i=1}^N z^3_i\right)\frac{\partial^3\phi(0)}{3!}}}}
+ \ldots \right)
e^{(\alpha_1+\alpha_2+bN)\phi(0)}:\ =
\nn\\
= q^{2\alpha_1\alpha_2} \prod_{i<j}^N (z_i-z_j)^{2b^2}
\prod_{i=1}^N z_i^{2b\alpha_1}(q-z_i)^{2b\alpha_2}\
:\left(1 +
\underline{
\frac{\alpha_2 q + b\sum_{i=1}^N z_i}{\alpha_1+\alpha_2+bN}\ L_{-1}
}
+\right.
\label{integrand001}
\ee
\be
+ \underline{\underline{
\frac{(4(\alpha_1+\alpha_2+bN)+2Q)\left(\alpha_2q+b\sum_{i=1}^Nz_i\right)^2
-\left(\alpha_2q^2+b\sum^N_{i=1}z_i^2\right)}
{2(\alpha_1+\alpha_2+bN)(4(\alpha_1+\alpha_2+bN)^2+2Q(\alpha_1+\alpha_2+bN)-1)}
\ L_{-1}^2 +}}
\\ \nn
\underline{\underline{
+2\frac{(\alpha_1+\alpha_2+bN)\left(\alpha_2q^2+b\sum_{i=1}^Nz_i^2\right)
-\left(\alpha_2q+b\sum_{i=1}^nz_i\right)^2}
{4(\alpha_1+\alpha_2+bN)^2+2Q(\alpha_1+\alpha_2+bN)-1}\ L_{-2}
}} +
\label{integrand002}
\ee
\be
+\underline{\underline{\underline{
\frac{AL^3_{-1}+2(\alpha_1+\alpha_2+bN)BL_{-1}L_{-2}+2(\alpha_1+\alpha_2+bN)CL_{-3}}
{6(\alpha_1+\alpha_2+bN)(4(\alpha_1+\alpha_2+bN)^2+2Q(\alpha_1+\alpha_2+bN)-1)((\alpha_1+\alpha_2+bN)^2+Q(\alpha_1+\alpha_2+bN)-1)}
}}}+
\\ \nn \left.
+\ldots \right)
e^{(\alpha_1+\alpha_2+bN)\phi(0)}:
\label{integrand003}
\ee
where
\be \nn
A=\left(\alpha_2q^3+b\sum\limits_{i=1}^Nz^3_i\right)
-(\alpha_1+\alpha_2+bN+Q)\left(\alpha_2q+b\sum\limits_{i=1}^Nz_i\right)
\left(\alpha_2q^2+b\sum\limits_{i=1}^Nz^2_i\right)+
\\ \nn
+\Big(4(\alpha_1+\alpha_2+bN)^2+6Q(\alpha_1+\alpha_2+bN)+2Q^2(\alpha_1+\alpha_2+bN)-2\Big)
\left(\alpha_2q+b\sum\limits_{i=1}^Nz_i\right)^3
\\ \nn
B=-4(\alpha_1+\alpha_2+bN)^2\left(\alpha_2q^3+b\sum\limits_{i=1}^Nz^3_i\right)
+\nn\\+4(\alpha_1+\alpha_2+bN)^2((\alpha_1+\alpha_2+bN)+Q)\left(\alpha_2q+b\sum\limits_{i=1}^Nz_i\right)
\left(\alpha_2q^2+b\sum\limits_{i=1}^Nz^2_i\right)+
\\ \nn
+4\Big(1-3(\alpha_1+\alpha_2+bN)^2-3Q(\alpha_1+\alpha_2+bN)\Big)
\left(\alpha_2q+b\sum\limits_{i=1}^Nz_i\right)^3
\ee
\be\nn
C=2(4(\alpha_1+\alpha_2+bN)^2+2Q(\alpha_1+\alpha_2+bN)-1)\Big(
(\alpha_1+\alpha_2+bN)^2\left(\alpha_2q^3+b\sum\limits_{i=1}^Nz^3_i\right)+
\\ \nn
-(\alpha_1+\alpha_2+bN)\left(\alpha_2q+b\sum\limits_{i=1}^Nz_i\right)
\left(\alpha_2q^2+b\sum\limits_{i=1}^Nz^2_i\right)
+2\left(\alpha_2q+b\sum\limits_{i=1}^Nz_i\right)^3\Big)
\ee
Underlined by one, two and three lines are the contributions
at levels one, two and three respectively.

\section{Structure constants $C_{\alpha_1\alpha_2}^{\,\alpha_1+\alpha_2+bN}$
from Selberg integrals. Level one}

Now, as suggested in \cite{MMSh1,MMSh2}, we
take integrals over $z_i$ along an open contour which
connects positions of the two original operators.
In order to perform integration at level one,
i.e. to integrate the first line in OPE,
(\ref{integrand001}), one needs the integrals
which are given by the now standard formulas
from ref.\cite{Selint}
(see also the Appendix in the present paper):
\be
q^{2\alpha_1\alpha_2}
\prod_{i=1}^N \int_0^q dz_i\, z_i^{2b\alpha_1}(q-z_i)^{2b\alpha_2}
\prod_{i<j}^N (z_i-z_j)^{2b^2} = \nn \\ =
q^{N+b^2N(N+1)}\cdot q^{2(\alpha_1+Nb)(\alpha_2+Nb)}
\prod_{i=1}^N \int_0^1 dz_i z_i^{2b\alpha_1}(1-z_i)^{2b\alpha_2}
\prod_{i<j}^N (z_i-z_j)^{2b^2} \ \stackrel{(\ref{IS0})}{=} \nn \\
= q^{N+b^2N(N+1)}\cdot q^{2(\alpha_1+Nb)(\alpha_2+Nb)}
\prod_{j=0}^{N-1} \frac{\Gamma\Big(1+2b\alpha_1+jb^2\Big)
\Gamma\Big(1+2b\alpha_2+jb^2\Big)\Gamma\Big(1+(j+1)b^2\Big)}
{\Gamma\Big(2+2b\alpha_1+2b\alpha_2+(j+N-1)b^2\Big)\Gamma\Big(1+b^2\Big)}
\label{integral000}
\ee
and
\be
q^{2\alpha_1\alpha_2}
\prod_{i=1}^N \int_0^q dz_i\, z_i^{2b\alpha_1}(q-z_i)^{2b\alpha_2}
\prod_{i<j}^N (z_i-z_j)^{2b^2}
\underline{\Big(q\alpha_2 + b\sum_{i=1}^N z_i\Big)} =\nn\\
= q^{(N+1)(1+b^2N)}\cdot q^{2(\alpha_1+Nb)(\alpha_2+Nb)}
\prod_{i=1}^N \int_0^1 dz_i z_i^{2b\alpha_1}(1-z_i)^{2b\alpha_2}
\prod_{i<j}^N (z_i-z_j)^{2b^2}
\Big(\alpha_2 + b\sum_{i=1}^N z_i\Big)
\ \stackrel{(\ref{IS0})\&(\ref{IS1})}{=} \nn \\
= q^{(N+1)(1+b^2N)}\cdot q^{2(\alpha_1+Nb)(\alpha_2+Nb)}
\prod_{j=0}^{N-1} \frac{\Gamma\Big(1+2b\alpha_1+jb^2\Big)
\Gamma\Big(1+2b\alpha_2+jb^2\Big)\Gamma\Big(1+(j+1)b^2\Big)}
{\Gamma\Big(2+2b\alpha_1+2b\alpha_2+(N-1+j)b^2\Big)\Gamma\Big(1+b^2\Big)}
\cdot\nn\\
\cdot\left(\alpha_2 + Nb\frac{1+2b\alpha_1+(N-1)b^2  }
{2+2b\alpha_1+2b\alpha_2+2(N-1)b^2}\right)
\label{integral001}
\ee
One can now extract the structure constants from the integrals
of (\ref{integrand001}). We do it first for $N=1$ where formulas are
just a little simpler, and then for arbitrary $N$.

$\bullet$ For $N=1$ these integrals are just the Euler $B$-functions:
\be
<1>\ =\int_0^1 dz\ z^{2b\alpha_1}(1-z)^{2b\alpha_2}
= \frac{\Gamma(1+2b\alpha_1)\Gamma(1+2b\alpha_2)}
{\Gamma(2+2b\alpha_1+2b\alpha_2)},\nn\\
<\alpha_2+bz>\ =
\int_0^1 dz\ z^{2b\alpha_1}(1-z)^{2b\alpha_2}\underline{(\alpha_2+bz)}
= \frac{\Gamma(1+2b\alpha_1)\Gamma(1+2b\alpha_2)}
{\Gamma(2+2b\alpha_1+2b\alpha_2)}
\left(\alpha_2 + b\frac{1+2b\alpha_1}{2+2b\alpha_1+2b\alpha_2}\right)
\label{lev1N1}
\ee
According to (\ref{OPEfs}) and (\ref{integrand001}),
the first of these formulas defines
\be
\tilde S_{\alpha_1\alpha_2}^{\alpha_1+\alpha_2+b}
\ \stackrel{(\ref{lev1N1})}{=}\ \frac{\Gamma(1+2b\alpha_1)\Gamma(1+2b\alpha_2)}
{\Gamma(2+2b\alpha_1+2b\alpha_2)},
\ee
while the second one is proportional to the product
$\tilde S_{\alpha_1\alpha_2}^{\alpha_1+\alpha_2+b}
\tilde C_{\alpha_1\alpha_2}^{\alpha_1+\alpha_2+b,L_{-1}}$.
Therefore, $\tilde C_{\alpha_1\alpha_2}^{\alpha_1+\alpha_2+b,L_{-1}}$
is given by the ratio of the two integrals (up to an additional factor):
\be
\tilde C_{\alpha_1\alpha_2}^{\alpha_1+\alpha_2+b,L_{-1}} =
\frac{<\alpha_2+bz>}{<1>}
\frac{1}{\alpha_1+\alpha_2+b} \ \stackrel{(\ref{lev1N1})}{=}\
\left(\alpha_2 + b\frac{1+2b\alpha_1}{2+2b\alpha_1+2b\alpha_2}\right)
\frac{1}{\alpha_1+\alpha_2+b}
\label{tildeCL-1}
\ee
At the same time,
from (\ref{Cfrefsc001}) in this case one has, taking into account that
$Q=b-1/b$:
\be
C_{\alpha_1\alpha_2}^{\alpha_1+\alpha_2+b,L_{-1}} =
\frac{\Delta_2+\Delta-\Delta_1}{2\Delta} \ \stackrel{(\ref{Delal})}{=}\
\frac{\alpha_2(\alpha_2-Q)+(\alpha_1+\alpha_2+b)(\alpha_1+\alpha_2+b-Q)
-\alpha_1(\alpha_1-Q)}{2(\alpha_1+\alpha_2+b)(\alpha_1+\alpha_2+b-Q)}
= \nn \\
= \frac{2\alpha_2(\alpha_1+\alpha_2)+2b\alpha_1+2\alpha_2/b+1}
{2(\alpha_1+\alpha_2+b)(\alpha_1+\alpha_2+1/b)}
= \left(\alpha_2 + b\frac{1+2b\alpha_1}{2+2b\alpha_1+2b\alpha_2}\right)
\frac{1}{\alpha_1+\alpha_2+b} = \nn \\
\stackrel{(\ref{lev1N1})}{=}\ \frac{<\alpha_2+bz>}{<1>}
\frac{1}{\alpha_1+\alpha_2+b} \ \stackrel{(\ref{tildeCL-1})}{=} \
\tilde C_{\alpha_1\alpha_2}^{\alpha_1+\alpha_2+b,L_{-1}}
\ee

$\bullet$ Similarly, for arbitrary $N$:
\be
C_{\alpha_1\alpha_2}^{\alpha_1+\alpha_2+bN,L_{-1}} =
\frac{\Delta_2+\Delta-\Delta_1}{2\Delta} =
\frac{\alpha_2(\alpha_2-Q)+(\alpha_1+\alpha_2+bN)(\alpha_1+\alpha_2+bN-Q)
-\alpha_1(\alpha_1-Q)}{2(\alpha_1+\alpha_2+bN)(\alpha_1+\alpha_2+bN-Q)}
= \nn \\
= \left(\alpha_2 + bN\frac{1+2b\alpha_1+(N-1)b^2}
{2+2b\alpha_1+2b\alpha_2+2(N-1)b^2}\right)
\frac{1}{\alpha_1+\alpha_2+bN} = \frac{<\alpha_2+ b\sum_i z_i>}{<1>}
\frac{1}{\alpha_1+\alpha_2+bN}
= \tilde C_{\alpha_1\alpha_2}^{\alpha_1+\alpha_2+bN,L_{-1}}
\ee
and
\be
\tilde S_{\alpha_1\alpha_2}^{\alpha_1+\alpha_2+bN}
\ \stackrel{(\ref{integral000})}{=}\
\prod_{j=0}^{N-1} \frac{\Gamma\Big(1+2b\alpha_1+jb^2\Big)
\Gamma\Big(1+2b\alpha_2+jb^2\Big)
}
{\Gamma\Big(2+2b\alpha_1+2b\alpha_2+(N-1+j)b^2\Big)
}
\prod_{j=1}^N \frac{\Gamma(1+jb^2)}{\Gamma(1+b^2)}
\ee
Formulas for the structure constants $\tilde C$ are rational,
therefore, they can be straightforwardly analytically continued in $N$
to arbitrary values of $\alpha = \alpha_1+\alpha_2+bN$.
The analytical continuation of the above expression for $\tilde S$ is somewhat
more ambiguous (and, anyway, there is nothing to compare them with, since
the coefficients $S$ generally do not factorize
into holomorphic and anti-holomorphic parts).

\section{Level two}

At level two, one needs to integrate (\ref{integrand002}).
The ordinary Selberg integrals (\ref{IS0}) and (\ref{IS1})
are not sufficient for this purpose, one needs also the generalized
one (\ref{IS2}) from the Appendix.
Then, the two integrals that one needs in (\ref{integrand002})
turn out to be
\be\nn
q^{2\alpha_1\alpha_2}
\prod_{i=1}^N \int_0^q dz_i\, z_i^{2b\alpha_1}(q-z_i)^{2b\alpha_2}
\prod_{i<j}^N (z_i-z_j)^{2b^2}
\underline{\underline{
\Big(\alpha_2q^2 + b\sum_{i=1}^N z^2_i\Big)
}} =
\\ \nn
= q^{(N+1)(1+b^2N)+1}\cdot q^{2(\alpha_1+Nb)(\alpha_2+Nb)}
\prod_{i=1}^N \int_0^1 dz_i z_i^{2b\alpha_1}(1-z_i)^{2b\alpha_2}
\prod_{i<j}^N (z_i-z_j)^{2b^2}
\Big(\alpha_2 + b\sum_{i=1}^N z^2_i\Big)
\ \stackrel{(\ref{IS0})\&(\ref{IS2})}{=} \\
= q^{(N+1)(1+b^2N)+1}\cdot q^{2(\alpha_1+Nb)(\alpha_2+Nb)}
\prod_{j=0}^{N-1} \frac{\Gamma\Big(1+2b\alpha_1+jb^2\Big)
\Gamma\Big(1+2b\alpha_2+jb^2\Big)\Gamma\Big(1+(j+1)b^2\Big)}
{\Gamma\Big(2+2b\alpha_1+2b\alpha_2+(N-1+j)b^2\Big)\Gamma\Big(1+b^2\Big)}
\left[\alpha_2+Nb\times\phantom{{a^2\over a^2}}\right.
\ee
{\footnotesize
\be
\left. \times\frac{
(4\alpha_1^2b^2+4\alpha_1\alpha_2b^2+6\alpha_1b^3N-8\alpha_1b^3+4b^3\alpha_2N-4b^3\alpha_2+8\alpha_1b+4\alpha_2b+4+3b^4N^2-7b^4N+4b^4+7b^2N-9b^2)
(2\alpha_1b+b^2N-b^2+1)}{2(2\alpha_1b+2\alpha_2b+2b^2N-3b^2+2)(2\alpha_1b+2\alpha_2b+2b^2N-2b^2+3)(\alpha_1b+\alpha_2b+b^2N-b^2+1)}
\right]\nn
\ee
}

\smallskip

\noindent
and
\be\nn
q^{2\alpha_1\alpha_2}
\prod_{i=1}^N \int_0^q dz_i\, z_i^{-2b\alpha_1}(q-z_i)^{-2b\alpha_2}
\prod_{i<j}^N (z_i-z_j)^{-2b^2}
\underline{\underline{
\Big(\alpha_2q + b\sum_{i=1}^N z_i\Big)^2
}} =
\\\nn
= q^{(N+1)(1+b^2N)+1}\cdot q^{2(\alpha_1+Nb)(\alpha_2+Nb)}
\prod_{i=1}^N \int_0^1 dz_i z_i^{2b\alpha_1}(1-z_i)^{2b\alpha_2}
\prod_{i<j}^N (z_i-z_j)^{2b^2}
\Big(\alpha_2 + b\sum_{i=1}^N z_i\Big)^2
\ \stackrel{(\ref{IS0})\&(\ref{IS2})}{=} \\
= q^{(N+1)(1+b^2N)+1}\cdot q^{2(\alpha_1+Nb)(\alpha_2+Nb)}
\prod_{j=0}^{N-1} \frac{\Gamma\Big(1+2b\alpha_1+jb^2\Big)
\Gamma\Big(1+2b\alpha_2+jb^2\Big)\Gamma\Big(1+(j+1)b^2\Big)}
{\Gamma\Big(2+2b\alpha_1+2b\alpha_2+(N-1+j)b^2\Big)\Gamma\Big(1+b^2\Big)}
\cdot
\ee
{\footnotesize
\be\nn
\cdot\left(\alpha_2^2 + 2N\alpha_2b\frac{1+2b\alpha_1+(N-1)b^2}
{2+2b\alpha_1+2b\alpha_2+2(N-1)b^2}
+N(N-1)b^2\frac{(2\alpha_1b+b^2N-b^2+1)(2\alpha_1b+b^2N-2b^2+1)}{2(\alpha_1b+\alpha_2b+b^2N-b^2+1)(2\alpha_1b+2\alpha_2b+2b^2N-3b^2+2)}+
Nb^2\times\right. \\
\left.\times\frac{
(4\alpha_1^2b^2+4\alpha_1\alpha_2b^2+6\alpha_1b^3N-8\alpha_1b^3+4b^3\alpha_2N-4b^3\alpha_2+8\alpha_1b+4\alpha_2b+4+3b^4N^2-7b^4N+4b^4+7b^2N-9b^2)
(2\alpha_1b+b^2N-b^2+1)}{2(2\alpha_1b+2\alpha_2b+2b^2N-3b^2+2)(2\alpha_1b+2\alpha_2b+2b^2N-2b^2+3)(\alpha_1b+\alpha_2b+b^2N-b^2+1)}
\right)\nn
\ee
}

\section{Higher levels}

The detailed explicit formulas are quite lengthy already at level two.
Writing them down for higher levels is simply impossible:
they take several pages.
However, in every particular case eq.(\ref{OPEfs}) can be easily
validated by simple computer calculations, provided one knows
the following set of matrices.

$\bullet$
The action of Virasoro generators on the free field primaries,
\be
L_{-Y} e^{\alpha\phi} = \sum_{|Y'|=|Y|} {\cal L}_{YY'}(\alpha)
:J^{Y'}e^{\alpha\phi}:
\ee
where $J^Y = \p^{n_1}\phi\,\p^{n_2}\phi\ldots$, produces
the matrix ${\cal L}_{YY'}$ (see s.2).

$\bullet$
The expansion
\be
e^{\alpha_1\phi(0) + \alpha_2\phi(q) + b\sum_i\phi(z_i)} =
\sum_Y {\cal E}_Y(q,\vec z) J^Y(0)
e^{(\alpha_1+ \alpha_2 + bN)\phi(0)}
\ee
gives the vector ${\cal E}_Y$ actually expressed through
the Schur polynomials.
Up to level 3 this vector, and also
\be
{\cal E}^Y = \sum_{|Y'|=|Y|}{\cal L}^{YY'} {\cal E}_{Y'}
\ee
with ${\cal L}^{YY'}$ being the inverse of ${\cal L}_{YY'}$,
are explicitly given in eqs.(\ref{integrand001})-(\ref{integrand003}).

$\bullet$
If $Y_1$ and $Y_2$ in (\ref{OPEfs}) are non-trivial, then
one actually needs a more sophisticated triple-vertex
${\cal E}_Y^{Y_1Y_2}$, describing the expansion
\be
:J^{Y_1}e^{\alpha_1\phi(0)}:\ :J^{Y_2}e^{\alpha_2\phi(q)}:\
:e^{b\sum_i\phi(z_i)}:=
\sum_Y {\cal E}_Y^{Y_1Y_2}(q,\vec z) :J^Y(0)
e^{(\alpha_1+ \alpha_2 + bN)\phi(0)}:
\ee
In this case, one has to consider also the quantity
\be
{\cal E}^{Y_1Y_2;Y} =
\sum_{Y_1',Y_2',Y'}
{\cal L}_{Y_1Y_1'}{\cal L}_{Y_2Y_2'}{\cal E}^{Y_1'Y_2'}_{Y'}
{\cal L}^{YY'}
\ee

$\bullet$
These ${\cal E}_Y$ are actually functions of $\{z_i\}$,
i.e. have the form
\be
{\cal E}_Y = \sum_{Y'}\hat{\cal E}_{YY'}\, z^{Y'}
\ee
with one extra index $Y'$. This time the sizes of Young diagrams
can be different, only $|Y'|\leq |Y|$.

$\bullet$
Integration over $z$ converts $\hat{\cal E}$ into
\be
<{\cal E}_Y>\ = \sum_{Y'} \hat{\cal E}_{YY'} I^Y
\ee
where $I^Y =\ <z^Y>\ $ are generalized Selberg integrals, described in the
Appendix below.
In fact, as emphasized in \cite{IO}, they are expressed
through the simpler quantities, the averages of Jack polynomials
with the help of one more matrix,
\be
I^Y = \sum_{|Y'|=|Y|} {\cal P}^{YY'}\!\! <P_{Y'}>
\ee
inverse to the matrix of expansion of the Jack polynomials
into monomials,
\be
P_Y = \sum_{Y'} {\cal P}_{YY'}z^{Y'}
\ee

\bigskip

Putting all the things together, one obtains for the
Dotsenko-Fateev representation
of the conformal triple function:
\fr{
\tilde C^Y_{Y_1Y_2} = {\cal L}_{Y_1Y_1'}(\alpha_1){\cal L}_{Y_2Y_2'}(\alpha_2)
{\cal L}^{Y'Y}(\alpha_1+\alpha_2+bN)
<{\cal E}^{Y_1'Y_2'}_{Y'}>,
\\
\\
<{\cal E}^{Y_1'Y_2'}_{Y'}> = \hat{\cal E}^{Y_1'Y_2'}_{Y'|Y''}
<z^{Y''}>, \ \ \ \ \ \ \
<z^{Y''}>\ = {\cal P}^{Y''Y'''}<P^{Y'''}>}
Summation over repeated indices is implied.

This should be compared with the usual CFT expression
\be
C^Y_{Y_1Y_2} = \sum_{Y'}\bar\Gamma_{Y_1Y_2Y'}Q^{YY'}
\ee
Details of this calculation are described in \cite{MMMM}.

An explicit check of the relation
\be
\boxed{
\tilde C^Y_{Y_1Y_2} = C^Y_{Y_1Y_2}
}
\label{C=C}
\ee
at levels 1 and 2 and for $Y_1=Y_2=\emptyset$
is described above in ss.5-6.
Since all the matrices are explicitly presented there also for
the case of level 3, it is a trivial computer exercise to make a check
also at this level, and, of course, it also confirms relation
(\ref{C=C}).

To check it at other levels for $Y_1=Y_2=\emptyset$
one needs to know four matrices,
${\cal L}$, $\hat{\cal E}$, ${\cal P}$, $Q$,
and two vectors, $<P^Y>$ and $\bar\Gamma_Y$.
When $Y_1$ and $Y_2$ are non-trivial,
$\hat{\cal E}$ and $\bar\Gamma$ acquire an additional
pair of indices, $Y_1,Y_2$.
These entries belong to different sciences:
${\cal L}$ and ${\cal E}$ to the free field calculus,
$Q$ and $\bar\Gamma$ to CFT,
${\cal P}$ to the theory of orthogonal polynomials,
$<P_Y>$ to the theory of Selberg integrals\footnote{Note that
the Selberg integrals (see the Appendix)
produce in the denominators the products
automatically presenting the decomposition of the Kac determinants
in terms of $\alpha$-variables, \cite{DF,CFT2}.
}.
Eq.(\ref{C=C}), the weak (matrix model) form of the AGT
conjecture establishes a concrete relation between
the seemingly unrelated quantities from these different subjects.

\section{Virasoro intertwiners}
Expansion rule (\ref{OPE}) implies that the structure constants
$C$ are the components of the Virasoro intertwining operator between
Verma modules $\Delta_1$, $\Delta_2$ and $\Delta$.
Comultiplication in the Virasoro algebra is somewhat non-trivial
\cite{MoSe}:
\be
\ldots \nn \\
\Delta(L_{-1}) = L_{-1}\otimes I + I\otimes L_{-1}, \nn \\
\Delta(L_{-0}) = L_{0}\otimes I + I\otimes L_{0}
+ q\ I\otimes L_{-1}, \nn \\
\Delta(L_{1}) = L_{1}\otimes I + I\otimes L_{-1}
+ 2q\ I\otimes L_0 + q^2\ I\otimes L_{-1}, \nn \\
\ldots
\label{vircom}
\ee
As the simplest example, this means that $L_0$ acts
on the operator product expansion of two primaries as
\be
L_0\Big(V_{\Delta_1}(0)V_{\Delta_2}(q)\Big) =
\left(\Delta_1+\Delta_2 + q\frac{\p}{\p q}\right)
\Big(V_{\Delta_1}(0)V_{\Delta_2}(q)\Big)\
\ee
This is, of course, in a perfect agreement with (\ref{OPE}):
\be
V_{\Delta_1}(0)V_{\Delta_2}(q) = q^{\Delta-\Delta_1-\Delta_2}
\sum_Y q^{|Y|}C^{\Delta,Y}{\Delta_1\Delta_2}L_{-Y}V_\Delta(0)
\ee
On one hand,
$\left(\Delta_1+\Delta_2 + q\frac{\p}{\p q}\right)
q^{\Delta-\Delta_1-\Delta_2+|Y|} = (\Delta+|Y|)
q^{\Delta-\Delta_1-\Delta_2+|Y|},$
on the other hand, $L_0L_{-Y}V_\Delta = (\Delta+|Y|)L_{-Y}V_\Delta$.
It is instructive to see how this works also for
representation (\ref{OPEfs}).
Then, at the l.h.s., one has a product of many operators,
$V_{\Delta_1}(0)V_{\Delta_2}(q)\left(\int_0^q V_1(z)dz\right)^N$,
and the multiple comultiplication (it is associative) now acts as
\be
L_0\otimes I\otimes I^{\otimes N}
+ I\otimes (L_0+qL_{-1})\otimes I^{\otimes N}
+ I\otimes I\otimes\Big((L_0+z_1L_{-1})\otimes I^{\otimes(N-1)} +\ldots
+ I^{\otimes(N-1)}\otimes (L_0+z_NL_{-1})\Big)
\ee
Since $V_1 = \ :e^{b\phi}:\ $ has unit dimension,
$(L_0+zL_{-1})V_1(z) = \frac{\p}{\p z}\big(zV_1(z)\big)$.
Our definition of Selberg integrals is the analytical continuation
from the points where all $2\alpha_ib$ and $b^2$ are positive
integers, therefore, all $z$-derivatives are always integrated
to zero as if one uses the {\it closed} contours, so that one actually gets
\be
L_0\left\{
V_{\Delta_1}(0)V_{\Delta_2}(q)\left(\int_0^q V_1(z)dz\right)^N
\right\}
= \left.\left(\Delta_1+\Delta_2+q\frac{\p}{\p q}\right)
\left\{V_{\Delta_1}(0)V_{\Delta_2}(q)\left(\int_0^{q'} V_1(z)dz\right)^N
\right\}\right|_{q'=q}
\label{L0DF}
\ee
Note that the $q'$-derivative does {\it not} act on the upper limit of
the integrals.
The multiple integral at the r.h.s. of this formula depends on $q$
through the factors $q^{2\alpha_1\alpha_2}$
and $\prod_{i=1}^N(q-z_i)^{2\alpha_2b}$
in the integrand.
The action of the logarithmic $q$-derivative on the first factor
gives the factor $2\alpha_1\alpha_2$, while the action on the others gives
$-2\alpha_2b \sum_{i=1}^N \frac{q}{q-z_i}$.
When $N=1$ this simply means that the Selberg (Euler) integral
has its argument $c=2\alpha_2b$ shifted by $-1$:
\be
2\alpha_2bq\int_0^q z^{2\alpha_1b}(1-z)^{2\alpha_2b-1}dz =
(2\alpha_1+2\alpha_2b+1)\int_0^qz^{2\alpha_1b}(1-z)^{2\alpha_2b}dz
\ee
For $N>1$ this is a similar, but a little more complicated exercise
(see eq.(\ref{1/z}) below for a similar evaluation of
$\left<\sum_i \frac{1}{z_i}\right>$),
which gives
\be
2\alpha_2b \left<\sum_{i=1}^N \frac{1}{q-z_i} \right> =
N\Big( 2\alpha_1b + 2\alpha_2b + (N-1)b^2+1\Big)
\ee
Substituting all this together with (\ref{OPEfs}) and
$\Delta_i = \alpha_i(\alpha_i-b+1/b)$
into the r.h.s. of (\ref{L0DF}),
one obtains for the coefficient in front of $V_\Delta$
\be
\alpha_1^2+\alpha_2^2-(\alpha_1+\alpha_2)(b-1/b) +
2\alpha_1\alpha_2
+ N\Big(2\alpha_1 b + 2\alpha_2b + (N-1)b^2+1\Big) = \nn \\
= (\alpha_1+\alpha_2+bN)(\alpha_1+\alpha_2+bN - b+1/b) = \Delta,
\ee
as needed.
In a similar way, one can act with any other $L_k$ on (\ref{OPEfs})
and check in detail that it is indeed consistent with the
comultiplication rule (\ref{vircom}).
This, of course, follows from the general argument
of \cite{DF}, since the screening insertion is an integral
of the dimension one operator, and, once again, our definition of
Selberg integrals actually allows one to consider the integration
contour as closed.
As we demonstrated in this section,
an explicit check confirms this general claim.

\section{From OPE to conformal blocks
\label{confb}}

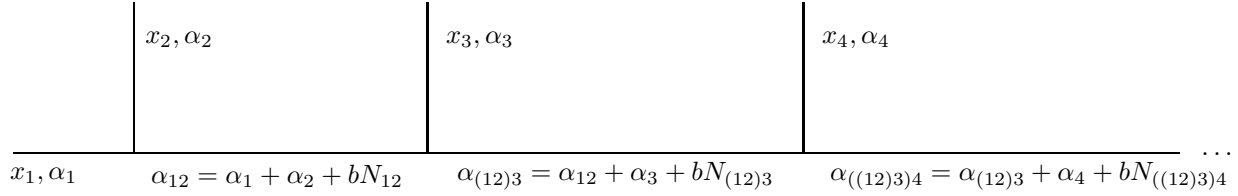
\begin{figure}
\unitlength 1mm 
\linethickness{0.4pt}
\ifx\plotpoint\undefined\newsavebox{\plotpoint}\fi
\begin{picture}(160,35)(-7,-7)
\put(-5,0){\line(1,0){155}}
\put(11,0){\line(0,1){20}}
\put(50,0){\line(0,1){20}}
\put(100,0){\line(0,1){20}}
\put(155,0){\makebox(0,0)[cc]{$\ldots$}}
\put(-1,-3){\makebox(0,0)[cc]{$x_1, \alpha_1$}}
\put(17,15){\makebox(0,0)[cc]{$x_2, \alpha_2$}}
\put(57,15){\makebox(0,0)[cc]{$x_3, \alpha_3$}}
\put(107,15){\makebox(0,0)[cc]{$x_4, \alpha_4$}}
\put(30,-3){\makebox(0,0)[cc]{$\alpha_{12}=\alpha_1+\alpha_2+bN_{12}$}}
\put(75,-3){\makebox(0,0)[cc]{$\alpha_{(12)3}=\alpha_{12}+\alpha_3+bN_{(12)3}$}}
\put(130,-3){\makebox(0,0)[cc]{$\alpha_{((12)3)4}
=\alpha_{(12)3}+\alpha_4+bN_{((12)3)4}$}}
\end{picture}
\caption{{\footnotesize
A comb-like conformal block from \cite{AGT,AlMo2},
for which the AGT relation is known in the case of
$0=x_1\ll x_2\ll x_3\ll x_4\ll \ldots$
Shown are the $\alpha$-parameters, the dimensions are equal to
$\Delta = \alpha(\alpha-b+1/b)$.
The intermediate dimensions are parameterized by the $N$-variables,
after analytical continuation in $N$ they take arbitrary
(continuum) values.
}}
\label{picCB1}
\end{figure}

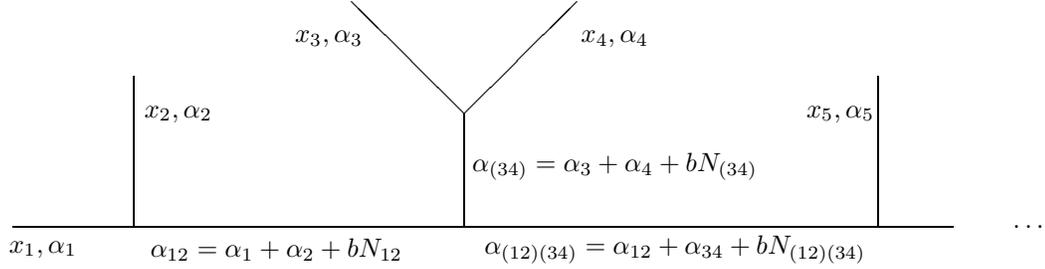
\begin{figure}
\unitlength 1mm 
\linethickness{0.4pt}
\ifx\plotpoint\undefined\newsavebox{\plotpoint}\fi
\begin{picture}(160,35)(-27,-7)
\put(-5,0){\line(1,0){125}}
\put(11,0){\line(0,1){20}}
\put(55,0){\line(0,1){15}}
\put(55,15){\line(-1,1){15}}
\put(55,15){\line(1,1){15}}
\put(110,0){\line(0,1){20}}
\put(130,0){\makebox(0,0)[cc]{$\ldots$}}
\put(-1,-3){\makebox(0,0)[cc]{$x_1, \alpha_1$}}
\put(17,15){\makebox(0,0)[cc]{$x_2, \alpha_2$}}
\put(37,25){\makebox(0,0)[cc]{$x_3, \alpha_3$}}
\put(75,25){\makebox(0,0)[cc]{$x_4, \alpha_4$}}
\put(105,15){\makebox(0,0)[cc]{$x_5, \alpha_5$}}
\put(30,-3){\makebox(0,0)[cc]{$\alpha_{12}=\alpha_1+\alpha_2+bN_{12}$}}
\put(75,8){\makebox(0,0)[cc]{$\alpha_{(34)}=\alpha_{3}+\alpha_4+bN_{(34)}$}}
\put(83,-3){\makebox(0,0)[cc]{$\alpha_{(12)(34)}
      =\alpha_{12}+\alpha_{34}+bN_{(12)(34)}$}}
\end{picture}
\caption{{\footnotesize
A star-like conformal block,
for which the AGT relation (the corresponding set of Nekrasov
function) is yet unknown.
The "matrix-model" or Dotsenko-Fateev representation,
a weak form of the AGT relation, is readily available
and provided by eqs.(\ref{CB2}) and (\ref{OPEfs})
of the present paper.
The brackets in subscripts of the $N$-variables define
the integration segments of the screening insertions.
From the point of view of the conformal blocks, they refer
to a particular corner of the $x$-moduli space, see \cite{AlMo2}.
}}
\label{picCB2}
\end{figure}

Eq.(\ref{OPEfs}) is very well suited for constructing arbitrary
conformal blocks. If we denote the operator product expansion at the l.h.s.
of Eq.(\ref{OPEfs}) through
$V_1(0)*_N V_2(q)$, then conformal block is the value of a linear form
on an ordered product, for example,
\be
\left<   \left(\Big(V_1(x_1)*_{N_{12}}V_2(x_2)\Big) *_{N_{(12)3}}V_3(x_3)\right)
*_{N_{((12)3)4}} V_4(x_4) *\ldots \right>
\label{CB1}
\ee
for Fig.\ref{picCB1} or
\be
\left<  \left(\Big(V_1(x_1)*_{N_{12}}V_2(x_2)\Big) *_{N_{(12)(34)}}
\Big(V_3(x_3)*_{N_{34}}V_4(x_4)\Big)\right) *_{N_{((12)(34))5}}
V_5(x_5)*\ldots\right>
\label{CB2}
\ee
for Fig.\ref{picCB2},
analytically continued in all the $N$-variables,
which are in this way converted into arbitrary intermediate dimensions.

The linear form here is defined by the usual rule
\be
\left< L_{-Y}e^{\alpha\phi(x)} \right>\ \sim \delta_{Y,\emptyset}
\delta_{\alpha,Q}
\ee

Note that the product $*_N$ is defined in (\ref{OPEfs}) asymmetrically:
the result is an operator at point $0$, i.e. at the position of the
first entry of the product. This makes $*_N$ non-associative:
\be
\Big(V_1(x_1)*_{N}V_2(x_2)\Big)*_M V_3(x_3) \equiv
V_1(x_1)V_2(x_2)V_3(x_3) \left(\int_{x_1}^{x_2} :e^{b\phi}:\right)^N
 \left(\int_{x_1}^{x_3} :e^{b\phi}:\right)^M
\ee
while
\be
V_1(x_1)*_{N}\Big(V_2(x_2)*_M V_3(x_3)\Big) \equiv
V_1(x_1)V_2(x_2)V_3(x_3) \left(\int_{x_1}^{x_2} :e^{b\phi}:\right)^N
\left(\int_{x_2}^{x_3} :e^{b\phi}:\right)^M
\ee
and the difference is in the integration segments in the last items.
Thus, the brackets are essential in the above expressions for
the conformal blocks.
In practice, CFT calculations are determined by the ordering of
$x$-arguments:
$0=x_1\ll x_2\ll x_3\ll x_4\ll \ldots $ in (\ref{CB1}) and
$0=x_1\ll x_2 \ll x_3 \ll x_5$, $x_4-x_3\ll x_3$
in (\ref{CB2}) and, hence, these two cases correspond to different
regions of the values of variables in the conformal block.

\section{Towards a proof of the AGT conjecture
\label{AGTproof}}

As explained in \cite{MMSh1,MMSh2} and further developed in \cite{IO},
representations like (\ref{CB1})
can be directly used to prove the original AGT conjecture.
For example, for the 4-point conformal block (\ref{CB1}) implies that
\be\label{4cb}
B = q^{\alpha_1\alpha_2}(1-q)^{\alpha_2\alpha_3}
\prod_{k=1}^{N_{12}} \int_0^q y_k^a (1-y_k)^c (y_k-q)^\gamma dy_k
\prod_{i=1}^{N_{(12)3}} \int_0^1 z_i^a (1-z_i)^c(z_i-q)^\gamma dz_i
\prod_{i<j} z_{ij}^{2\beta} \prod_{k<l} y_{kl}^{2\beta}
\prod_{i,k} (z_i-y_k)^{2\beta}
\ee
where $q = \frac{x_{21}x_{34}}{x_{24}x_{31}}$, $x_1=0$, $x_2=q$, $x_3=1$,
$N_1=N_{12}$,
$N_2=N_{(12)3}$ and $a = 2b\alpha_1$, $\beta=b^2$, $c=2b\alpha_3$,
$\gamma=2b\alpha_2$.
In order to make use of the Selberg integrals from the Appendix,
which are all along the segment $[0,1]$, we rescale $y_k\rightarrow qy_k$
and expand in powers of $q$:
\be
B = q^{{\rm deg}(B)}
\prod_{i=1}^{N_{(12)3}}\int_0^1 z_i^{a+\gamma +2\beta N_1}(1-z_i)^c dz_i
\prod_{i<j} z_{ij}^{2\beta}\ \
\prod_{k=1}^{N_{12}} \int_0^1
y_k^a (1-y_k)^\gamma dy_k \prod_{k<l} y_{kl}^{2\beta}
\cdot\nn \\
\cdot \exp\left\{-2\sum_{m=1}^\infty \frac{q^m}{m}
\left(\alpha_2 + b \sum_{i=1}^{N_{(12)3}}\frac{1}{z_i^m}\right)
\left(\alpha_3 + b \sum_{k=1}^{N_{12}} y_k^m\right)
\right\}
\ee
This nice exponential formula first appeared in \cite{IO}.

It can be now expanded in the Schur/Jack polynomials so that the result is
a bilinear combination of the generalized Selberg integrals over the $z$ and $y$
variables. The integrals are labeled by Young diagrams (see the Appendix below),
thus, one naturally obtains a bilinear expansion in Young diagrams.
In the formulation of \cite{IO}, the AGT conjecture is now reduced to
the claim that there are two different expansions of this type:
in triple vertices and in the Nekrasov functions.
Denoting independent averaging over the $z$ and $y$ variables by the double
angle brackets, we have \cite{IO}:
\be
\left<\left< \exp\left\{-2\sum_{m=1}^\infty \frac{q^m}{m}
\left(\alpha_2 + b \sum_{i=1}^{N_{(12)3}}\frac{1}{z_i^m}\right)
\left(\alpha_3 + b \sum_{k=1}^{N_{12}} y_k^m\right)
\right\}\right>\right>
\begin{array}{l}
\longrightarrow\sum_{Y_1,Y_2} \bar\Gamma(Y_1)Q^{-1}(Y_1,Y_2) \Gamma(Y_2)
\cr
\cr
\cr\longrightarrow
(1-q)^{2\alpha_2\alpha_3} \sum_{Y_1,Y_2}Z_{Nek}(Y_1,Y_2)
\end{array}
\label{nekex}
\ee

\bigskip

$\bullet$
Despite this is already done in \cite{MMSh1,MMSh2,IO},
for the sake of completeness
we explicitly illustrate the situation at the first level of the $q$-expansion.

At this level, one needs just two explicit expressions for the Selberg integrals
from the Appendix:
\be
<\sum y>\ = \frac{N_{12}I[1]}{I[0]}
= N_{12}\frac{a+(N_{12}-1)\beta +1}{a+\gamma+(2N_{12}-2)\beta +2}
\ee
and
\be\label{1/z}
\left<\sum \frac{1}{z}\right>\ = \frac{N_{(12)3}I[-1]}{I[0]}
= \frac{N_{(12)3}I_{a'-1}[1^{N_{(12)3}-1}]}{I_{a'}[0]}=
N_{(12)3}\frac{a'+c+(N_{(12)3}-1)\beta +1}{a'}
\ee
Here $a' = a+\gamma+2N_1\beta$ and
\be
-\left< \alpha_2\alpha_3+2\beta \sum_i \frac{1}{z_i} \sum_k y_k
+ \gamma \sum_i \frac{1}{z_i} + c \sum_k y_k\right>\ =
\frac{(\Delta+\Delta_2-\Delta_1)(\Delta+\Delta_3-\Delta_4)}{2\Delta}
\ee
with \cite{MMSh2}
\be
\Delta = a'(a+\gamma+(2N_1-2)\beta +2)/2\beta = (a+\gamma+2N_1\beta)
(a+\gamma+2N_1\beta+2-2\beta)/2\beta,\nn \\
\Delta_1 = ( a)(a + 2-2\beta)/2\beta, \nn \\
\Delta_2 = ( \gamma )(\gamma + 2-2\beta)/2\beta, \nn \\
\Delta_3 = ( c )(c + 2-2\beta)/2\beta, \nn \\
\Delta_4 = ( a+c+\gamma+2\beta(N_1+N_2) )
(a+c+\gamma+2\beta(N_1+N_2) + 2-2\beta)/2\beta
\ee
thus reproducing the expression for the $q$-linear contribution to
the conformal block \cite{CFT1,CFT2,MMMagt}.

\bigskip

$\bullet$
In general the $q^m$-term of the $B$-expansion contains
bilinear combinations of the integrals $I_{N_1}[Y]$ and $I_{N_2}[-Y']$
with $|Y|,|Y'|\leq m$.
Generalization to the multi-point conformal blocks with the multi-linear
expansion in Young diagrams is also straightforward.

\section{Conclusion
\label{conc}}

In this paper we justified the claim that the coefficients
of the operator product expansions in arbitrary conformal theory
are fully controlled by the free field model,
provided one allows insertions of the Dotsenko-Fateev screening operators
between the points, where the original operators are located,
and analytical continuation in the number of these insertions.
The well-known complexity of the operator expansion coefficients
appears related to that of the generalized Selberg integrals,
which are defined for arbitrary Young diagrams, but
contain non-trivial non-factorizable polynomial factors
when the diagrams are different from $[1^n]$.
Since the single line diagrams $[1^n]$ are associated with
the hypergeometric series \cite{MMnf}, one may say that the non-triviality
of the Selberg integrals for other diagrams is responsible for the
deviation of the conformal blocks from the hypergeometric functions
and, thus, it is what requires the generic Nekrasov functions to appear
in description of the conformal blocks.

\subsection{The proof of the matrix-model version of AGT conjecture}

Despite the present paper does not contain a full {\it constructive} proof,
hopefully, it provides a conceptually clear explanation of the week form of
the AGT conjecture \cite{MMSh1,MMSh2}, identifying the conformal block
with the analytically continued matrix model partition function in the DV phase
\cite{DVagt,Egu,Ito,Wilma}.
Moreover, for the 3-point functions this identification can
be {\it implicitly} (not constructively) proved with the following
chain of arguments:

$\bullet$
The structure constants
$C_{\alpha_1,Y_1; \alpha_2,Y_2}^{\alpha,Y}$
in (\ref{OPE}), i.e. components of the intertwining operator,
are unambiguously defined by
representation theory of the Virasoro algebra.

$\bullet$
Free field + DF -- induced structure constants
$\tilde C_{\alpha_1,Y_1; \alpha_2,Y_2}^{a,Y}$
in (\ref{OPEfs}) are also components of the Virasoro intertwining
operator,
but for a triple of concrete and {\it explicitly} realized
Verma modules.
Instead they are defined only for discrete values of
$a = \alpha_1+\alpha_2+bN$.

$\bullet$
The both $C$ and $\tilde C$ are rational functions of their
arguments $\alpha$ and $a$.
The rational analytical continuation in $a$ of
the function $\tilde C$ is unique, therefore,
such an analytical continuation coincides with $C$:
\be
\boxed{
C_{\alpha_1,Y_1; \alpha_2,Y_2}^{\alpha,Y} =
\tilde C_{\alpha_1,Y_1; \alpha_2,Y_2}^{\alpha,Y}
}
\ee

\bigskip

Note that with this technique one obtains "matrix-model"
representations for arbitrary conformal blocks, not only for (\ref{CB1}),
but also for (\ref{CB2}).
At the same time, the literal AGT relations are currently applicable only
for the case of (\ref{CB1}), their generalization (an extension
of the set of the Nekrasov functions) to arbitrary conformal blocks
remains to be found.

\subsection{Towards a proof of the remaining part of the AGT conjecture}

After the matrix model version of the AGT conjecture is proved, the
original AGT conjecture is reduced to an exercise, outlined above in
section \ref{AGTproof}.
There are still some combinatorial identities to be proved in
this direction, but this seems rather straightforward.
{\bf F complete proof of the AGT conjecture on this track is now clearly
within reach.}

Much more interesting would be to prove the AGT conjecture
differently, without any direct use of the Nekrasov functions.
Given the result of the present paper, it turns into
a puzzling observation that the two kinds of integrals are identical:
the matrix model (Dotsenko-Fateev) integrals like (\ref{4cb})
and the LNS ones \cite{LNS} like
\be
\sum_k {q^k\over k!}\left({\epsilon\over\epsilon_1\epsilon_2}\right)^k
\prod_{a=1}^k \int{dx_a\over 2\pi i}\frac{\prod_{i=1}^4(x_a+m_i)}{(x_a^2-
(\alpha_{12}-\epsilon/2)^2)
((x_a+\epsilon)^2-(\alpha_{12}-\epsilon/2)^2)}
\prod_{a<b}^k
\frac{x_{ab}^2(x_{ab}^2-\epsilon^2)}{(x_{ab}^2-\epsilon_1^2)(x_{ab}^2-\epsilon_2^2)}
\label{LNS}
\ee
where $\alpha_{12}=\alpha_1+\alpha_2+bN_{12}$, $\epsilon=\epsilon_1
+\epsilon_2$, $\epsilon_1=b$, $\epsilon_2=-1/b$ and the four parameters $m_i$,
$i=1...4$ are linear combinations of $\alpha_{1,2,3}$ and $\alpha_4\equiv
\alpha_{12}+\alpha_3+bN_{(12)3}$ \cite{MMMagt}. Thus, the positions of poles
in the LNS integral (dictated by $\alpha_{12}$ and, hence, by $N_{12}$)
become a number of integrations in the DF case, while the number
of integrations in the LNS integral depends on the degree of expansion into
$q$, i.e. on the level of expansion of the conformal block into
descendant contributions.

Thus, the AGT acquires form of a duality relation, where the number of
integrations
on one side is a parameter in the integrand at the other side and vice versa.
This type of duality may seem mysterious, but it is well-known in
the theory of matrix models \cite{UFN3}.
The simplest example is provided by conversion of the Gaussian model
into the Kontsevich type model \cite{GaKo}:
\be
\int_{N\times N} dM e^{-{1\over 2}\Tr M^2 + \sum_k t_k\Tr M^k}
\sim \int_{n\times n} dX (\det X)^N e^{-{1\over 2}\tr X^2 -i\tr \Lambda X}
\label{GaKo}
\ee
where $\Tr,{\rm Det}$ and $\tr,\det$ denote the traces
and determinants of the $N\times N$ and $n\times n$ matrices,
and $t_k = \frac{1}{k}\tr \Lambda^{-k}$.
Both models are of the eigenvalue type and
clearly the number $N$ of integrations over the eigenvalues
at the l.h.s. appears just as a parameter at the r.h.s.,
where the number of integrations is a fully independent parameter $n$.
This identity, (\ref{GaKo}) can also be rewritten as
\be
\int_{N\times N} dM \ {\rm Det}\Big(I\otimes I - M\otimes \Lambda^{-1}\Big)
\ e^{-\Tr M^2}
\sim \int_{n\times n} dX\ \det\left(I - i\frac{X}{\Lambda}\right)^N e^{
-{1\over 2}\tr X^2}
\ee
which, if expanded in powers of $\Lambda^{-1}$, becomes an identity
for the Gaussian correlators:
\be
1 -\ \frac{1}{2}\tr\frac{1}{\Lambda^2} <<\Tr M^2>>_N +\
\frac{1}{2}\left(\tr\frac{1}{\Lambda}\right)^2\!\!
<<(\Tr M)^2>>_N + \ldots = \nn \\
= 1 + \frac{N}{2}\left<\tr X\frac{1}{\Lambda} X\frac{1}{\Lambda}\right>_n -
\frac{N^2}{2}\left<\left(\tr X\frac{1}{\Lambda}\right)^2\right>_n + \ldots
\ee
Since $<<M_{ij} M_{kl}>>\ = \delta_{jk}\delta_{il}$
and $<X_{ab}X_{cd}>\ = \delta_{bc}\delta_{ad}$, the two sides
of the equality coincide, but, as usual for dualities,
the second term at the l.h.s. is equal to the third term at the r.h.s.
and vice versa.

Eq.(\ref{GaKo}) can be used (at $\beta=1$) to further transform the multiple
integrals (\ref{4cb}) so that $N$ becomes a parameter in the integrand.
Indeed, such integrals arise from the l.h.s. of (\ref{GaKo}) for
$t_k = \sum_a \frac{\mu_a}{k}q_a^k$ which for the integer values of $\mu_a$
correspond to $\Lambda$ matrices of the block form
$\Lambda = \sum_a q_a I_{\mu a}$ at the r.h.s. of (\ref{GaKo}).
The number of integrations at the r.h.s. is then equal to $n = \sum_a \mu_a$.
It is natural to conclude that  {\bf the AGT relation
between (\ref{4cb}) and (\ref{LNS})
is a further generalization of the duality relation (\ref{GaKo})} continued to
the $\beta$-ensembles (to $\beta\neq 1$) and to non-integer values of $n$.
Details of this analysis will be presented elsewhere.

\subsection{Extension from Virasoro to $W$}

Another mystery associated with our result in this paper
concerns extension to the case of several free fields.
In conformal theory, this corresponds to switching from the Virasoro
to $W$ chiral algebras.
The problem is that representation theory of the $W$ algebras is {\it not}
sufficient to specify unambiguously arbitrary conformal blocks.
Additional constraints should therefore be imposed by brute force.
It is, however, unclear what are the parallel restrictions in
the free field formalism and its Dotsenko-Fateev extension
described in the present paper, which seems easily generalizable to
an arbitrary number of free fields.
This subject also remains open for future investigation.

\section*{Acknowledgements}

Al.Mor. is indebted for the hospitality and support
to Uppsala University, where part of this work was done.

Our work is partly supported by Russian Federal Nuclear Energy Agency,
Federal Agency for Science and Innovations of
Russian Federation under contract
02.740.11.5194, by RFBR grants 10-01-00536
(A.Mir. and Al.Mor.) and 10-01-00836 (An.Mor.),
by joint grants 09-02-90493-Ukr,
09-02-93105-CNRSL, 09-01-92440-CE, 09-02-91005-ANF, 10-02-92109-Yaf-a.

\newpage

\section*{Appendix. Selberg integrals and their generalizations}

The Selberg integrals
\be
I_Y =
\prod_{i=1}^N\int_0^q dz_i \left\{z^{Y} \prod_{i<j}^N (z_i-z_j)^{2\beta}
\prod_{i=1}^N z_i^{a}(q-z_i)^{c}\right\}
\label{Selin}
\ee
with $z^Y = z_1^{n_1}z_2^{n_2}\ldots$ for $Y = \{n_1\geq n_2\geq\ldots\}$
are direct generalizations of the Euler Beta-function,
also represented as products of elementary Gamma-function factors.
The Selberg integrals are naturally labeled by Young diagrams $Y$,
and well-known are only integrals for the single line diagrams $[1^n]$.
For more complicated diagrams, the integrals contain additional polynomial
factors, which are not further factorized into linear expressions.
However, they are needed for comparison with DF 3-point functions
in this paper.

$\bullet$
From \cite{Selint} one knows that
\be
I{[0]} = \prod_{j=1}^{N} \frac{\Gamma(\beta j+1)}{\Gamma(\beta+1)}
\prod_{j=0}^{N-1}
\frac{\Gamma(a+\beta j+1)\Gamma(c+\beta j+1)}{\Gamma\Big(a+c+(N-1+j)\beta+2\Big)}
\label{IS0}
\ee
\be
I{[1]} = \frac{a + (N-1)\beta+1}{a+c+(2N-2)\beta + 2}\ I{[0]}
\label{IS1}
\ee
and, in general,
\be
I{[1^n]} = I{[0]}\ \prod_{j=1}^n \frac{a+(N-j)\beta+1}{a+c+(2N-j-1)\beta+2}
\label{IS1n}
\ee

$\bullet$
If the
Young diagram $Y$ contains $k>1$ lines, then \cite{Selint} is not sufficient,
and actually $I[Y]$ acquires
additional factors, which are polynomials of degree $2k-2$.
In particular,
\be
I[2] = \frac{a^2+ac + (3N-4)a\beta  + 2(N-1)\beta c + 4a + 2c + 4 + (N-1)(3N-4)\beta ^2 +
(7N-9)\beta }{\Big(a+c+(2N-3)\beta +2\Big)\Big(a+c+(2N-2)\beta  + 3\Big)}\ I[1]
 = \nn \\
= \frac{\Big(a+(N-1)\beta +1\Big)
\Big(a^2+ac + (3N-4)a\beta  + 2(N-1)\beta c  + (N-1)(3N-4)\beta ^2
+ 4a + 2c + (7N-9)\beta  + 4\Big)
}{\Big(a+c+(2N-3)\beta +2\Big)\Big(a+c+(2N-2)\beta  +2\Big)
\Big(a+c+(2N-2)\beta  + 3\Big)}\ I[0]
\label{IS2}
\ee
\be
I[21] =
\frac{\Big(a+(N-2)\beta+1\Big)\Big(a+(N-1)\beta+1\Big)\ I[0]}
{\Big(a+c+(2N-4)\beta +2\Big)\Big(a+c+(2N-3)\beta +2\Big)
\Big(a+c+(2N-2)\beta +2\Big)\Big(a+c+(2N-2)\beta +3\Big)}\cdot\nn \\
\cdot \Big(a^2+ac+(3N-5)a\beta  + (2N-3)\beta c+3(N-1)(N-2)\beta ^2 + 4a+2c
+ (7N-12)\beta  + 4\Big)
\label{IS21}
\ee
and in general
\be
I[21^n] = I[0]\
\left(\frac{a^2+ac+(3N-4)(N-1)\beta^2 + (3N-4)a\beta + 2(N-1)c\beta
+4a+2c+(7N-9)\beta +4}{a+c+(2N-2)\beta + 3} - n\beta\right)\cdot \nn \\
\cdot\frac{
\prod_{j=1}^{n+1} \Big(a+(N-j)\beta + 1\Big)}
{\prod_{j=1}^{n+2} \Big(a+c+(2N-j-1)\beta + 2\Big)}
\ee
Moving further,
{\footnotesize
\be
I[3]=\frac{\Big(a+(N-1)\beta +1\Big)\ I[0]\ P[3]}
{\Big(a+c+(2N-4)\beta +2\Big)
\Big(a+c+(2N-3)\beta +2\Big)\Big(a+c+(2N-2)\beta +2\Big)
\Big(a+c+(2N-2)\beta +3\Big)\Big(a+c+(2N-2)\beta +4\Big)}
\label{IS3}
\ee
}
where
$$
P[3] = (a+3)(a+2)(a+c-\beta +2)(a+c-2\beta +2)
+$$ $$+ (N-1)\beta\Big((a+c+2)(6a^2+5ac+32a+11c+40
+ 4\beta^2) + 2\beta(a+2c+2)(2a+c+5)\Big)
+ $$ $$+ (N-1)(N-2)\beta^2\Big((16a^2+21ac+5c^2 + 80a+52c+98)
+2\beta(20a+16c+51) + 36\beta^2\Big)
+ $$ $$ + 2(N-1)(N-2)(N-3)\beta^3 (10a+24\beta+7c+26)
\ +\ 10(N-1)(N-2)(N-3)(N-4)\beta^4
$$

\bigskip

$\bullet$
One can get rid of the non-factorizable polynomials in $I[Y]$
by switching to peculiar linear combinations.
For example,
\be
I[2] +\frac{(N-1)\beta}{1+\beta} I[11] =
\frac{1+N\beta}{1+\beta}
\frac{\Big(a+(N-1)\beta +1\Big)\Big(a+(N-1)\beta +2\Big)}
{
\Big(a+c+(2N-2)\beta  +2\Big)
\Big(a+c+(2N-2)\beta  + 3\Big)}\ I[0]
\label{2li11}
\ee
Note that not only a decomposition into linear factors
is obtained in this way, also
the factor $\Big(a+c+(2N-3)\beta +2\Big)$, which was present in the
denominators of both $I[2]$ and $I[11]$, is canceled in this combination.

As noted in \cite{IO} the relevant linear combinations are
actually the Jack polynomials $P^{(1/\beta)}[Y]$:
\be
P^{(1/\beta)}[1^n](z) = m_{(1^n)}(z)=\sum_{1\le i_1<i_2<\ldots<i_n}
\prod_{k=1}^n z_{i_k}, \nn \\
P^{(1/\beta)}[2] = m_2(z)+{2\beta\over 1+\beta}m_{(1^2)}(z)=
\sum_i z_i^2+{2\beta\over 1+\beta}\sum_{1\le i<j}z_iz_j, \nn \\
P^{(1/\beta)}[3] =m_{[3]}(z)+{3\beta\over 1+2\beta}m_{[2,1]}(z)+
{6\beta^2\over (1+\beta)(1+2\beta)}m_{[1^3]}(z),\ \ \ \ \ \
P^{(1/\beta)}[2,1] =m_{[2,1]}(z)+ {6\beta\over 1+2\beta}m_{(1^3]}(z),\nn \\
\ldots
\ee
Indeed, according to \cite{Kad}, the Selberg integrals of Jack polynomials
are factorized:
\be
<P^{(1/\beta)}[Y]> \ = c[Y] I[0]\ \prod_{i\geq 1} \prod_{j=0}^{n_i-1}
\frac{a + (N-i)\beta + 1+j}{a+c+(2N-1-i)\beta+2+j}
\ee
and the $\beta$- and $N$-dependent coefficient is
\be
c[Y] = \frac{\prod_{i\geq 1} \prod_{j=0}^{n_i-1} (N+1-i)\beta+j }
{\prod_{(i,j)\in Y} \Big(n_i-j + (\tilde n_j-i+1)\beta\Big)}
\ee
where $\tilde n$ parameterizes the transposed Young diagram
$\tilde Y = \{\tilde n_1\geq \tilde n_2\geq\ldots\}$.
In particular,
\be
c[2] = \frac{N\beta(N\beta+1)}{(n_1-1+\tilde n_1\beta)(n_1-2+\tilde n_2\beta)}
= \frac{N\beta+1}{\beta+1},
\ee
in accordance with (\ref{2li11}).

\bigskip

$\bullet$
The Selberg integrals $I[Y]$ satisfy a set of sum rules.

\noindent
Since $\prod_i^N (1-z_i) = 1 - \sum_{i=1}^N z_i + \sum_{i<j}^N z_iz_j - \ldots\ $
one has
\be
I_{c+1}[0] = I_c[0] - NI_c[1] + \frac{N(N-1)}{2}\,I_c[11] - \ldots
= \sum_{n=0}^N \frac{(-)^nN!}{n!(N-n)!}\, I_c[1^n]
\ee
what is indeed true for (\ref{IS1n}).
This sum rule involves only the single row Young diagrams.

Similarly, from the expansion $\prod_i^N (1-z_i)^2 = 1 - 2\sum_{i=1}^N z_i +
\sum_{i=1}^N z_i^2 + 2\sum_{i<j}^N z_iz_j - \ldots\ $ one gets
\be
I_{c+2}[0] = I_c[0] - 2NI_c[1] + NI_c[2] + N(N-1)I_c[11] - \ldots
\ee
which includes only the double row Young diagrams
(of which the single row diagram is a particular case with $k_2=0$).

Similarly, expanding $\prod_i^N (1-z_i)^m$, one can deduce
the expansion of $I_{c+m}[0]$ into a sum of the $m'$-row Young
diagrams with $m'\leq m$.
Moreover, such sum rules can also be written for $I_{c+m}[Y]$
with arbitrary $Y$.

\end{document}